\documentclass[12pt,preprint]{aastex}

\usepackage{graphicx}
\usepackage{natbib}
\usepackage[usenames,dvips]{color}
\usepackage[bookmarks=false]{hyperref}

\usepackage{amssymb}
\usepackage{epstopdf}
\slugcomment{Not to appear in Nonlearned J., 45.}

\shorttitle{Emission from Inner Disk and Corona} \shortauthors{Erlin
Qiao \& B.F. Liu}

\begin{document}


\title{The Emission from Inner Disk and Corona in
the Low and Intermediate Spectral States of Black Hole X-ray Binaries}


\author{Erlin Qiao\altaffilmark{1} and B. F.  Liu \altaffilmark{1}}
\affil{National Astronomical Observatories, Chinese Academy of
Sciences, Beijing 100012, China}

\email{qiaoel@nao.cas.cn}

\altaffiltext{1}{National Astronomical Observatories, Chinese
Academy of Sciences, Beijing 100012, China}

\begin{abstract}
Recent observations reveal that a cool disk may survive in the
innermost stable circular orbit (ISCO) for some black hole X-ray
binaries in the canonical low/hard state. The spectrum is
characterized by a power law with a photon index $\Gamma \sim
1.5-2.1$ in the range of 2-10 keV and a weak disk component with
temperature of $\sim 0.2$ keV. In this work, We revisit the
formation of such a cool, optically thick, geometrically thin disk
in the most inner region of black hole X-ray binaries at the
low/hard state within the context of disk accretion fed by
condensation of hot corona. By taking into account the cooling
process associated with both Compton and conductive processes in a
corona, and the irradiation of the hot corona to the disk, we
calculate the structure of the corona. For viscosity parameter
$\alpha=0.2$, it's found that the inner disk can exist for accretion
rate ranging from $\dot M \sim 0.006-0.03 \dot M_{\rm Edd}$, over
which the electron temperatures of the corona are in the range of
$1-5\times 10^9\ \rm K$ producing the hard X-ray emission. We
calculate the emergent spectra of the inner disk and corona for
different mass accretion rates. The effect of viscosity parameter
$\alpha$ and albedo $a$ ($a$ is defined as the energy ratio of
reflected radiation from the surface of the thin disk to incident
radiation upon it from the corona) to the emergent spectra are also
presented. Our model is used to explain the recent observations of
GX 339-4 and Cyg X-1, in which the thin disk may exist at ISCO
region in the low/hard state at luminosity around a few percent of
$L_{\rm Edd}$. It's found that the observed maximal effective
temperature of the thermal component and the hard X-ray photon index
$\Gamma$ can be matched well by our model.
\end{abstract}


\keywords{Accretion, accretion disks
--- Black hole physics
--- X-rays: individual (GX 339-4, Cyg X-1)
--- X-rays: stars}

\section{Introduction}
Black hole X-ray binaries (BHXRBs) are binary star systems that are
luminous in the X-ray part of the spectrum. These X-ray emissions
are generally thought to be caused by one of the component stars
being a black hole accreting matter from the companion. The property
of these systems have been reviewed by several authors (Remillard \&
McClintock 2006; Done et al. 2007; Gilfanov et al. 2010). It is
well-known that various spectral states have been exhibited in
BHXRBs. In particular, two basic X-ray spectral states are presented
with a high/soft spectral state occurring at high luminosities and a
low/hard spectral state occurring at low luminosities. Generally, at
the high/soft state, with higher accretion rate, the accretion is
dominantly via a standard thin disk extending to the innermost
stable circular orbit (ISCO) (Pringle \& Rees 1972; Shakura \&
Sunyaev 1973; Mitsuda et al. 1984; Frank et el. 2002); The ISCO is
at $3R_{\rm S}$ (where $R_ {\rm S}=2GM/c^2$, G is gravitational
constant, c is light speed, and $M$ is the central black hole mass)
for non-rotating black holes and $0.615 R_{\rm S}$ for rotating
black hole with limiting spin rate $a_{\ast}=0.9982$, where
$a_{\ast}$ is specific angular momentum (Thorne 1974); With decline
of the accretion rate to some transition rate, BHXRBs enter the
low/hard spectral state, in which the standard thin disk is replaced
by a hot, optically thin, geometrically thick advection dominated
accretion flows (ADAF) in the most inner region around the black
hole. i.e., the thin disk truncates at some radius off the ISCO
(Rees et al. 1982; Narayan \& Yi 1994, 1995a, b; Abramowicz et al.
1995; Narayan 2005, 2008; Kato et al. 2008; Esin et al. 1997;
Kawabata \& Mineshige 2010; Qiao et al. 2010; Zhang et al. 2010).

However, the picture of truncated outer thin disk + inner ADAF model
in the low/hard state is challenged by the recent observations of
some BHXRBs. Reis et al. (2010) investigate a sample composed of
eight BHXRBs in the low/hard state observed by {\em XMM-Newton} and
{\em Suzaku}. Although the hard X-ray continuum is characterized by
a power law with a photon index in the range of $\Gamma \sim
1.5-2.1$, a thermal component with a color temperature consistent
with $L\propto T^4$ is detected in all eight sources, meanwhile,
broad iron $\rm K_{\rm \alpha}$ fluorescence line is also observed
in half of the sample. Both the $L\propto T^4$ relation and the
broad iron line profile suggest that a cool disk extends to ISCO.
The observed thermal component is around $0.2 \rm \ keV$.
Observations to GX 339-4 are made by {\em Swift} and {\em RXTE} at
2007 May when the transition to the low/hard state occurs (Kalemci
et al. 2007). Fits to the observed broad iron $\rm K_{\alpha}$ line
profile with relativistic reflection model at luminosity $0.8\%
L_{\rm Edd}$ suggests that a cool disk resides in the region very
close to ISCO. A thermal component with a disk temperature $\sim
0.165\rm \ keV$ at luminosity $0.8\% L_{\rm Edd}$ is detected. The
X-ray spectra are roughly characterized by a power law with a photon
index $\Gamma=1.63^{+0.04}_{-0.03}$ (Tomsick et al. 2008).

A disk accretion model maintained by the recondensation of hot
corona has been proposed to understand the presence of the cool disk
in the most inner region around the black hole in the low/hard state
by Liu et al. (2006), Meyer et al. (2007). In this model, when the
accretion rate decreases just below the transition rate between
high/soft state and low/hard state, as a consequence of efficient
evaporation, the thin disk truncates at a distance where the maximum
evaporation rate occurs (Meyer et al. 2000; Liu et al. 1999, 2002).
The remnant disk from the truncation radius inwards can exist
steadily fed by the condensation of the hot corona/ADAF rather than
be swallowed by the black hole within the viscous time scale (Liu et
al. 2007; Taam et al. 2008; Liu et al. 2011). The size of the inner
remnant disk is governed by the accretion rate. The geometry of the
accretion flows are allocated as an inner disk and a much cooler
outer disk, which are separated by an ADAF. With decrease of the
accretion rate, the inner disk shrinks and eventually vanishes
completely at a certain accretion rate. The dynamical interaction
between the inner disk and corona is studied by Liu et al. (2006),
Meyer et al. (2007) in which only the vertical conductive cooling of
the corona is considered throughout the corona. Furthermore, the
Compton cooling of the corona by the soft photos from the underlying
thin disk is taken into account. The cooling of the corona is
classified as Compton-dominated case and conduction-dominated case
at different radius. For Compton-dominated case, conductive cooling
is neglected, and for conduction-dominated case, Compton cooling is
neglected (Liu et al. 2007). However, for a certain mass accretion
rate, Compton cooling and conductive cooling are comparable. In this
case, both the Compton cooling and conductive cooling should be
considered together. The cooling of the corona is modified by Taam
et al. (2008), in which the effect of conductive cooling to the
corona is added to Compton-dominated case, the effect of Compton
cooling to the corona is added to conduction-dominated case.

In this paper, in order to get a radially continuous,
self-consistent disk-corona model, we employ the result of Taam et
al. (2008) and consider the irradiation of the hot corona to the
disk (Liu et al. 2011). We calculate the structure of the corona and
the emergent spectra of the inner disk and corona with different
parameters of the model (In our calculation, contribution of the
ADAF gas in the gap between the inner disk and outer disk, and the
outer disk and corona to the spectra are not included). For the
emission of inner disk, both the accretion luminosity contributed
from disk accretion and the irradiation of the corona are included.
Our model is used to explain the recent observations of black hole
X-ray binaries GX 339-4 and Cyg X-1 in the low/hard spectral state.
In section 2, the disk-corona model is briefly described. In Section
3, we show the numerical results of the structure and the emergent
spectra of the inner disk-corona system. In Section 4, we compare
the model predictions with observations. Our conclusions are
presented in Section 5.

\section{The model}\label{model}
The disk-corona model adopted here, is based on the study of Taam et
al. (2008) and Liu et al. (2011). It is assumed that an ADAF-like
hot corona (described by the self-similar solution of Narayan \& Yi
1995, with $\alpha=0.2$, $\beta=0.8$) lies above a thin disk. The
corona is heated by the viscous release of gravitational energy of
accreted gas and cooled by vertical conduction and inverse Compton
scattering of soft photons emitted by the underlying disk. For the
Compton cooling of the corona, the upward hard photons escape from
the corona directly; the downward hard photons are partially
reflected and partially absorbed by the underlying optically thick
disk. The absorbed photons are reprocessed in the optically thick
disk and remitted as a thermal emission providing the soft photons
for the Compton scattering in the corona. This procedure is iterated
until a stable disk-corona system forms. In the vertical transition
layer between the disk and corona, an equilibrium is established
between the conductive flux from the upper corona, bremsstrahlung
radiation, and vertical enthalpy flux. For a given distance from the
black hole, a fraction of the disk gas is heated and evaporated to
the corona when the conduction flux is too large to be radiated
away. On the other hand, a certain amount of coronal gas is cooled
down, condensing to the disk if the bremsstrahlung radiation is more
efficient than the conduction. At accretion rates around a few
percent of Eddington value, gas evaporates from the disk to the
corona, the disk vanishes at around a few hundred Schwarzschild
radii, and the coronal gas partially condenses back to the disk in
the innermost region.

The cooling of the corona is classified as conduction-dominated case
and Compton-dominated case at different radius (Liu et al. 2007).
Here, we follow the work of Taam et al. (2008) in which the
conduction-dominated case is modified by adding the Compton cooling,
and the Compton-dominated case is modified by adding the conductive
cooling. Meanwhile, in our calculation the irradiation of the corona
to disk is added for getting a self-consistent disk-corona system
(Liu et al. 2011). Throughout the paper, we scale some quantities
as, $m$ is the central black hole mass scaled with solar mass
$M_{\odot}$. $\dot m$ is the accretion rate scaled with Eddington
accretion rate $\dot M_{\rm Edd}$, $\dot M_{\rm Edd}= 1.39\times
10^{18}m \,{\rm g \ s^{-1}}$. $r$ is the distance from the black
hole scaled with Schwarzschild radius $R_{\rm S}$, $R_ {\rm
S}=2GM/c^2=2.95 \times 10^5 m \, {\rm cm}$. For the sake of clarity,
we list the basic results of the disk-corona model in the following.

\subsection{Corona Dominated by Conductive Cooling} In the case of
conduction dominated cooling the condensation rate is
given as (Taam et al. 2008) \\
\begin{equation}\label{cnd1}
\dot m_{\rm cnd}(x)=3.23 \times 10^{-3} \alpha^{-7}\dot m^3 f(x),
\end{equation}

\begin{eqnarray}\label{conduction}
f(x)=\bigg({r_o\over r_1}\bigg)^{3/5}\bigg[6 \bigg( {r_1 \over
r_o}\bigg)^{1/10}-6\bigg({r_1 \over r_o}\bigg)^{1/10}x^{1/2}-3
\int_{r_i /r_o}^1 (1+\lambda)^{-2/5}x^{-2/5} dx\bigg],
\end{eqnarray}
where $\lambda={q_{\rm Cmp} \over dF_{\rm c}/dz}$ is the ratio of
Compton cooling rate to conductive cooling rate, which weighs the
modification of neglected cooling process compared to the dominated
cooling process. The ratio can be reexpressed as
$\lambda=1.4052\times 10^4 m r^{-3/2}\bigg[1-\bigg({3\over
r}\bigg)\bigg] T_{\rm eff,max}^4$, where $T_{\rm eff,max}$ is the
maximum effective temperature of the accretion disk (Taam et al.
2008). $r_1$ and $r_o$ represent the size of the inner disk without
and with modification in conduction-dominated case respectively.
$r_1$ and $r_o$ are expressed as,
\begin{eqnarray}\label{cndr}
r_1&=&0.815\alpha^{-28/3}\dot m^{8/3}, \nonumber \\
r_o&=&0.815\alpha^{-28/3}\dot m^{8/3}[1+\lambda(r_o)]^4.
\end{eqnarray}

The electron temperature distribution of the corona in radial
direction is,
\begin{eqnarray}
T_{\rm ec}=2.01\times 10^{10}\alpha^{-2/5}\dot
m^{2/5}r^{-2/5}(1+\lambda)^{-2/5} \rm K.
\end{eqnarray}

The Bremsstrahlung luminosity from the transition layer is
\begin{eqnarray}\label{brem}
{L_{\rm brem} \over L_{\rm Edd}}=0.0642\alpha^{-7/3}\dot
m^{5/3}\bigg[1-\bigg({3 \over r_o}\bigg)^{1/2}\bigg],
\end{eqnarray}

and the Compton luminosity from the corona is
\begin{eqnarray}\label{Cmp1}
{L_{\rm Cmp} \over L_{\rm Edd}}=1.349 \alpha^{-7/5}m\dot
m^{7/5}\bigg({T_{\rm eff, max} \over 0.3 \rm \ keV}\bigg)^{4}
\nonumber \\
\times
\int_{r_i/3}^{r_o/3}(1+\lambda)^{-2/5}x^{-29/10}(1-x^{-1/2})dx.
\end{eqnarray}

Note that the emission from the outer pure ADAF is not included
here, it would cause the coronal luminosity deviating from the true
value when the inner disk is very small, which occurs at accretion
rates less than 0.01.

\subsection{Corona Dominated by Compton Cooling} In the case of
Compton dominated cooling the condensation rate is given as,

\begin{eqnarray}\label{cnd2}
\dot m_{\rm cnd}(x)&=&A\bigg\{2B\bigg[\bigg({r_o \over
r_i}\bigg)^{1/2}-1\bigg]-\int_{r_i /3}^{r_o /3} \bigg({1+{1 \over
\lambda}}\bigg)^{-2/5}x^{1/5}\big(1-x^{-1/2}\big)^{-2/5}dx \bigg \},
\end{eqnarray}
where
\begin{eqnarray}\label{ab}
A&=&6.164 \times 10^{-3}\alpha^{-7/5}m^{-2/5}\dot
m^{7/5}\bigg({T_{\rm eff, max} \over 0.3 \rm \ keV}\bigg)^{-8/5}, \nonumber \\
B&=&3.001\alpha^{-14/15}m^{2/5}\dot m^{4/15}\bigg({T_{\rm eff, max}
\over 0.3 \rm \ keV}\bigg)^{8/5}\bigg({r \over 3}\bigg),
\end{eqnarray}
and the condensation radius $r_o$ meets the following formula,
\begin{eqnarray}\label{Cmpr}
r_o \bigg[1-\bigg({3 \over
r_o}\bigg)^{1/2}\bigg]^{-4/7}=14.417\alpha^{-4/3}m^{4/7}\dot
m^{8/21}  \nonumber \\
\bigg({T_{\rm eff, max} \over 0.3 \rm \ keV}\bigg)^{16/7}\bigg({1+{1
\over \lambda(r_o)}}\bigg)^{4/7}.
\end{eqnarray}

The electron temperature distribution of the corona in radial
direction is,
\begin{eqnarray}
T_{\rm em}=3.025\times 10^9\alpha^{-2/5}m^{-2/5}\dot m^{2/5} r^{1/5}
\nonumber \\
\times \bigg[1-\bigg({3\over
r}\bigg)^{1/2}\bigg]^{-2/5}\bigg(1+{1\over \lambda} \bigg)^{-2/5}
\bigg({T_{\rm eff,max} \over 0.3 \rm \ keV}\bigg)^{-8/5}.
\end{eqnarray}

The expression of Bremsstrahlung luminosity from the transition
layer is same as equation (\ref{brem}), and the the Compton
luminosity from the corona is,
\begin{eqnarray}\label{Cmp2}
{L_{\rm Cmp} \over L_{\rm Edd}}=0.392 \alpha^{-7/5}m^{3/5}\dot
m^{7/5}\bigg({T_{\rm eff, max} \over 0.3 \rm \ keV}\bigg)^{12/5}
\nonumber \\
\times \int_{r_i/3}^{r_o/3}(1+ {1\over
\lambda})^{-2/5}x^{-23/10}(1-x^{-1/2})dx.
\end{eqnarray}

For given black hole mass $m$, accretion rate $\dot m$, viscosity
parameter $\alpha$, and a presumed value of $T_{\rm eff, max}$,
quantities describing the inner disk and corona are determined by
eqs. (\ref{cnd1}) to (\ref{Cmp1}) in the case of conduction
dominated cooling or by (\ref{cnd2}) to (\ref{Cmp2}) in the case of
Compton dominated cooling. Iterative calculations are carried out
until a self-consistent $T_{\rm eff, max}$ is obtained.

\subsection{Iterative Calculations with Irradiation} The surface
effective temperature of the accretion disk with irradiation from
the corona is expressed as,
\begin{eqnarray}\label{T}
\sigma T_{\rm eff}^4(r)=F_{\rm cnd}+F_{\rm irr},
\end{eqnarray}
where $F_{\rm cnd}$ refers to the flux which originates from disk
accretion fed by condensation in per unit area, i.e.,
\begin{eqnarray}\label{Fcnd}
F_{\rm cnd}={3GM\dot M_{\rm cnd} \over 8\pi
R^3}\bigg[1-\bigg({3R_{\rm S}\over R}\bigg)^{1/2}\bigg],
\end{eqnarray}
$F_{\rm irr}$ refers to the the illumination flux from the corona to
the disk surface per unit area (Liu et al. 2011),
\begin{eqnarray}\label{Firr}
F_{\rm irr}={1 \over 2} L_{\rm c,in}(1-a) {H_{\rm s} \over
{4\pi(R^2+H_{\rm
s}^2)^{3/2}}} \nonumber \\
L_{\rm c,in}=L_{\rm brem}+L_{\rm Cmp},
\end{eqnarray}
where the ADAF-like corona is assumed above the disk as a point
source at a height $H_{\rm s}$, and the covering factor of the point
source to a disk ring at distance $R$ is given as $f={H_{\rm s}
\over {4\pi(R^2+H_{\rm s}^2)^{3/2}}}$, $a$ is albedo. $L_{\rm c,in}$
is the total intrinsic luminosity of the corona and the transition
layer above and below the disk.

Combining eqs. (\ref{T})-(\ref{Firr}) the effective temperature of
the accretion disk can be re-expressed as,
\begin{eqnarray}\label{Trp}
T_{\rm eff}(r)= 2.05T_{\rm eff,max}^{\prime} \bigg({3 \over r}
\bigg)^{3/4}
\bigg[1-\bigg({3 \over r} \bigg)^{1/2} \bigg]^{1/4} \nonumber \\
\times \bigg[{1+6L_{\rm c,in}(1-a) \over \dot M_{\rm cnd}c^2}
{H_{\rm s} \over 3R_{\rm s}}\bigg]^{1/4} \nonumber \\
=2.05T_{\rm eff,max} \bigg({3 \over r} \bigg)^{3/4} \bigg[1-\bigg({3
\over r} \bigg)^{1/2} \bigg]^{1/4},
\end{eqnarray}
where
\begin{eqnarray}\label{tmaxp}
T_{\rm eff,max}=T_{\rm eff,max}^{\prime} \bigg[{1+6L_{\rm c,in}(1-a)
\over \dot M_{\rm cnd}c^2} {H_{\rm s} \over 3R_{\rm s}}\bigg]^{1/4}.
\end{eqnarray}
$T_{\rm eff,max}^{\prime}$ refers to the maximum effective
temperature from disk accretion which is reached at $r_{\rm
tmax}=(49/12)$. The expression of $T_{\rm eff,max}^{\prime}$ is
given as (Liu et al. 2007),
\begin{eqnarray}\label{tmax}
T_{\rm eff, max}^{\prime}=0.2046 \bigg({m\over 10}
\bigg)^{-1/4}\bigg[{\dot m_{\rm cnd(r_{\rm tmax})}\over
0.01}\bigg]^{1/4} \ \rm keV.
\end{eqnarray}

At lower accretion rate $\dot m$, it's assumed that the corona is
dominated by conductive cooling. An effective temperature $T_{\rm
eff,max}$ is presumed to calculate the condensation rate from eqs.
(\ref{cnd1}) and (\ref{conduction}), the luminosity of the
transition layer and the corona from eqs. (\ref{brem}) and
(\ref{Cmp1}), with which a new effective temperature $T_{\rm
eff,max}$ is derived from eqs. (\ref{tmaxp}) and (\ref{tmax}) by
assuming a value of albedo $a$. An iteration is made till the
presumed temperature is consistent with the derived value. From the
derived effective temperature, the Compton-dominated region is
determined the following equation (Liu et al. 2007), which is,
\begin{equation}\label{r_cmp}
r_{\rm cmp}\bigg[{1-\bigg({3\over r_{\rm
cmp}}\bigg)^{1/2}}\bigg]^{-2/3}\le 23.487m^{2/3}\bigg({T_{\rm
eff,max}\over 0.3\ \rm keV}\bigg)^{8/3},
\end{equation}
If eq. (\ref{r_cmp}) has no solution, it means the corona is
dominated by conductive cooling throughout the corona, we find a
self-consistent solution of the disk-corona system. Otherwise, the
Compton-dominated region is determined by eq. (\ref{r_cmp}). We
recalculate the condensation rate by combing eqs. (\ref{cnd1}),
(\ref{conduction}) and (\ref{cnd2}), (\ref{ab}), and the luminosity
of transition layer and corona from eqs. (\ref{brem}), (\ref{Cmp1})
and (\ref{Cmp2}) till the presumed temperature is consistent with
the derived value, we find a solution of disk-corona system. With
increase of the mass accretion rate, Compton-dominated region
extends inward and outward till Compton cooling dominates throughout
the corona at some accretion rate. An effective temperature $T_{\rm
eff,max}$ is assumed to calculate the condensation rate from eqs.
(\ref{cnd2}), (\ref{ab}), and the luminosity from eqs. (\ref{brem})
and (\ref{Cmp2}), with which a new effective temperature $T_{\rm
eff,max}$ is calculated from eqs. (\ref{tmaxp}) and (\ref{tmax}).
Iterations are made till the presumed temperature is consistent with
the derived value, we find a solution of the disk-corona system.

\subsection{Calculation of the Spectra}
With determination of the disk and corona features, we are able to
calculate the spectrum of the disk and corona. The contribution to
the emergent spectra are composed of two components: (1) Multi
black-body emission of the underlying thin disk which is partially
scattered by the electrons in the hot corona (Compton emission) (2)
Bremsstrahlung emission of the transition layer between the disk and
corona. We use Monte Carlo simulation to calculate the Comptom
spectra from the hot corona. The method of the Monte Carlo
simulation is based on Pozdniakov, Sobol' \& Sunyaev (1977). We
assume the electrons in the corona have a Maxwellian distribution,
and the Compton scattering is the main cooling process. Since the
electron-scattering optical depth in the corona is less than one, we
introduce the weight $w$ as described by Pozdniakov et al. (1977) in
order to efficiently calculate the effects of multiple scattering.
We first set $w_0=1$ for a given soft photon, then calculate the
escape probability $P_0$ of passing through the slab. The quantity
of $w_0P_0$ are the transmitted portion and is recorded to calculate
the penetrated spectrum or reprocessed photons according to the
escape direction of the photon. The remaining weight
$w_1=w_0(1-P_0)$ is the portion that undergoes at least one
scattering. If we write the escape probability after the $n$-th
scattering as $P_n$, the quantity $w_nP_n$ is the transmitted
portion of photons after the $n$-th scattering, and is recorded as
upward or downward transmitted spectrum. The remaining portion
$w_n(1-P_n)$ undergoes the $(n+1)$-th scattering. This calculation
is continued until the weight $w$ becomes sufficiently small. The
whole process is simulated by the Monte Carlo method.

Because the temperature of the transition layer is obviously higher
than the underlying cool disk, we expect the inverse Compton
scattering of the photons from the transition layer in the hot
corona is not important Compared with the soft photons from the
underlying cool disk. So, we ignore the inverse Compton scattering
of the bremsstrahlung photons from the transition layer in the hot
corona. The bremsstrahlung spectral flux from the transition layer
per unit area can be expressed as,
\begin{eqnarray}\label{sp}
F_{\nu}^{\rm brem}&=& {1\over 2} j_{\nu}Z_{\rm cpl} \nonumber \\
&=&{1\over 2}\times 6.8\times 10^{-38}Z^2n_{e}n_{i}T^{-1/2}_{\rm
cpl}\overline{g_{\rm ff}} (\nu,T_{\rm cpl})e^{-{h \nu}/{kT_{\rm cpl}}}Z_{\rm cpl} \nonumber \\
&=&f\overline {g_{\rm ff}} e^{-{h \nu}/{kT_{\rm cpl}}} \rm \ ergs \
s^{-1} \ cm^{-2} Hz^{-1}.
\end{eqnarray}
where $j_{\rm \nu}$ refers to the volume emissivity coefficient
which is, $j_{\rm \nu}=6.8\times 10^{-38}Z^2n_{e}n_{i}T^{-1/2}_{\rm
cpl}$ $\overline{g_{\rm ff}} (\nu,T_{\rm cpl})e^{-{h \nu}/{kT_{\rm
cpl}}}$, and $f=3.4\times 10^{-38}Z^2n_{e}n_{i}T^{-1/2}_{\rm cpl}
Z_{\rm cpl}$. $\rm Z_{\rm cpl}$ is the thickness of the transition
layer, and $T_{\rm cpl}=1.98\times 10^9 \alpha^{-4/3}\dot m^{2/3}
\rm K$ is the electron temperature of the transition layer (Liu et
al. 2007). In eq. (19), the coefficient $1/2$ means that only the
upward photons from the transition layer can directly escape from
the disk-corona system. The frequency-integrated flux from the
transition layer is,
\begin{eqnarray}\label{flux}
F_{\rm brem}&=&\int_{0}^{\infty} F_{\rm \nu}^{\rm brem} d\nu=f
\int_{0}^{\infty} \overline{g_{\rm ff}} (\nu,T_{\rm cpl})e^{-{h
\nu}/{kT_{\rm cpl}}} d\nu \nonumber \\
&=&{fkT_{\rm cpl}\over h}\int_{0}^{\infty} \overline g_{\rm
ff}(\nu,T_{\rm cpl})e^{-x} dx \ \ \rm ergs \ s^{-1} \ cm^{-2}.
\end{eqnarray}
Solving eq. (\ref{flux}), we can get,
\begin{eqnarray}\label{f}
f={F_{\rm brem} h \over {kT_{\rm cpl}}}{1\over
\int_{0}^{\infty}\overline g_{\rm ff}(\nu,T_{\rm cpl})e^{-x} dx},
\end{eqnarray}
the expression of $F_{\rm brem}$ is given as (Liu et al. 2007),
\begin{eqnarray}
F_{\rm brem}={1\over 2}\times6.391\times
10^{24}\alpha^{-7/3}m^{-1}\dot m^{5/3}r^{-5/2}  \ \rm ergs \ s^{-1}
\ cm^{-2},
\end{eqnarray}
where the integral $\int_{0}^{\infty}\overline g_{\rm ff}(\nu,T_{\rm
cpl})e^{-x}dx$ is a frequency average of the velocity averaged Gaunt
factor, which is in the range 1.1 to 1.5. Choosing a value of 1.2
will give an accuracy to within about 12\% (Rybicki \& Lightman
1979).

By combing eq.(\ref{sp}) and eq. (\ref{f}), we get the local
emergent bremsstrahlung spectra from the transition layer, which is
given as,
\begin{eqnarray}
F_{\nu}^{\rm brem}&=&  {F_{\rm brem}h \over kT_{\rm cpl}} {1\over
\int_{0}^{\infty}\overline g_{\rm ff}(\nu,T_{\rm cpl})e^{-x}
dx}\overline {g_{\rm ff}}(\nu, T_{\rm cpl}) e^{-{h \nu}/{kT_{\rm
cpl}}} \rm \ \ \ \ ergs \ s^{-1} \ cm^{-2} Hz^{-1}.
\end{eqnarray}

By integrating the local Compton spectra from the corona and the
bremsstrahlung spectra from the transition layer along radial
direction, we get the total emergent spectra of the inner
disk-corona system.

\section{Numerical Results}\label{numerical}
Given a black holes mass $m$, the mass accretion rate $\dot m$, the
viscosity parameter $\alpha$ and the albedo $a$, the structure of
the corona in radial direction is solved self-consistently as
described in Section \ref{model}. Throughout the calculation, $m=10$
is adopted.

\begin{figure*}
\includegraphics[width=85mm,height=70mm,angle=0.0]{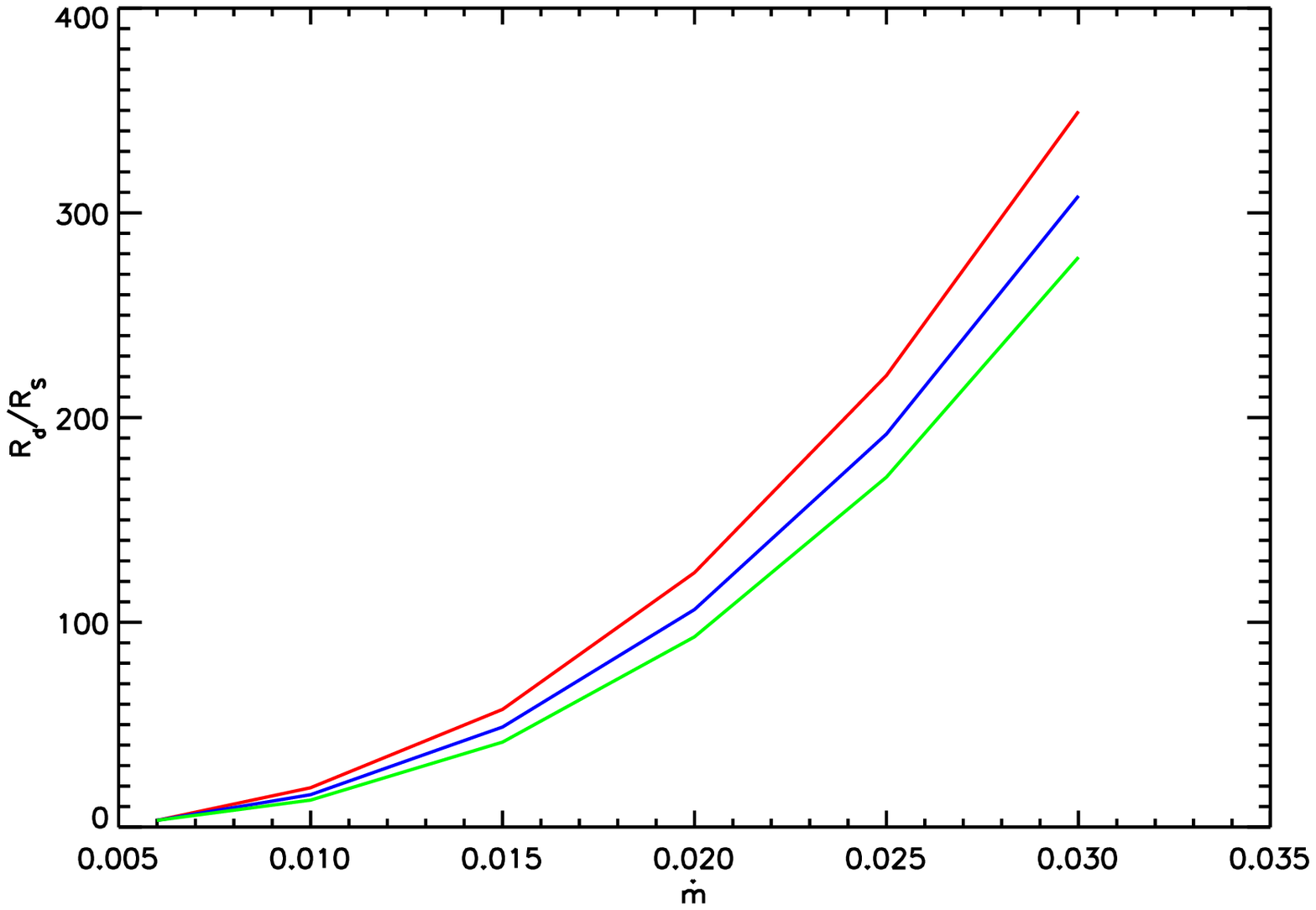}
\includegraphics[width=85mm,height=70mm,angle=0.0]{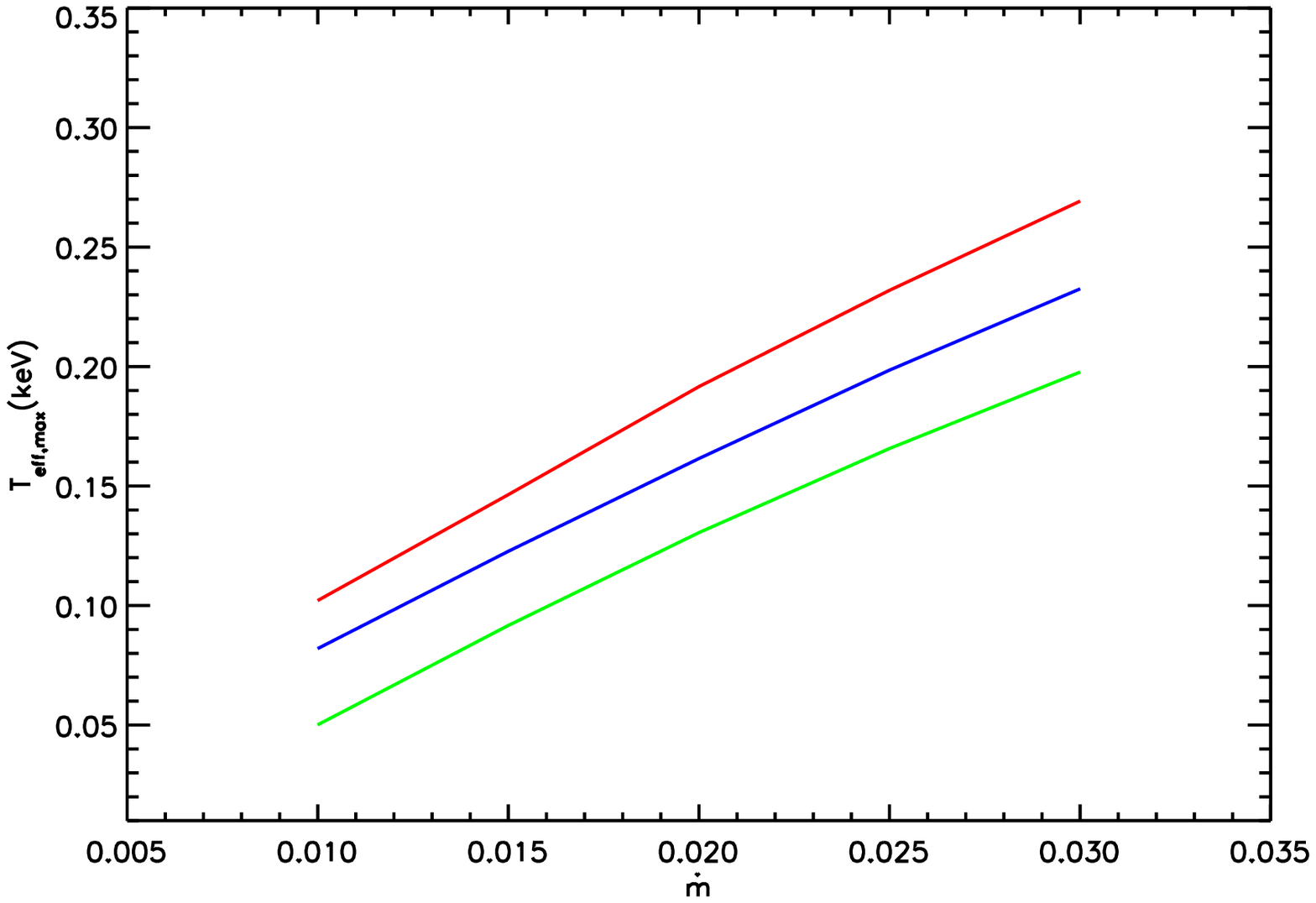}
\caption{\label{teff}The left panel: The size of the inner accretion
disk as functions of mass accretion rate $\dot m$. In our
calculation, $\alpha=0.2$ is adopted. The red line is for albedo
$a=0.15$ (blue line: $a=0.6$; green line: $a=1$). The right panel:
The maximum temperature of the disk as functions of mass accretion
rate $\dot m$.}
\end{figure*}

\begin{figure*}
\includegraphics[width=85mm,height=70mm,angle=0.0]{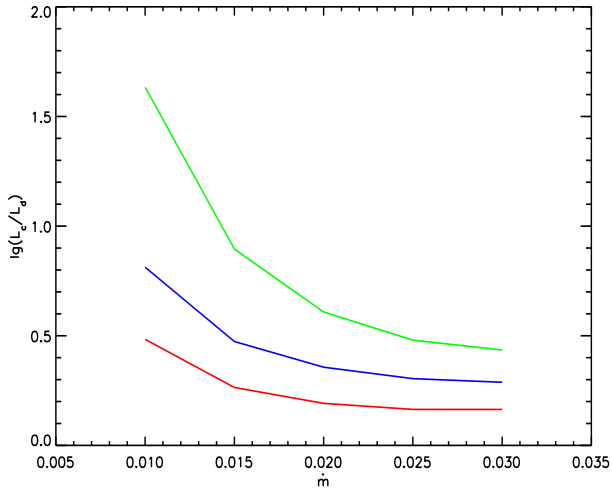}
\caption{\label{ratio}The ratio of $L_{\rm c}/L_{\rm d}$ as
functions of mass accretion rate $\dot m$. In our calculation,
$\alpha=0.2$ is adopted. The red line is for albedo $a=0.15$ (blue
line: $a=0.6$; green line: $a=1$).}
\end{figure*}

\begin{figure*}
\includegraphics[width=85mm,height=70mm,angle=0.0]{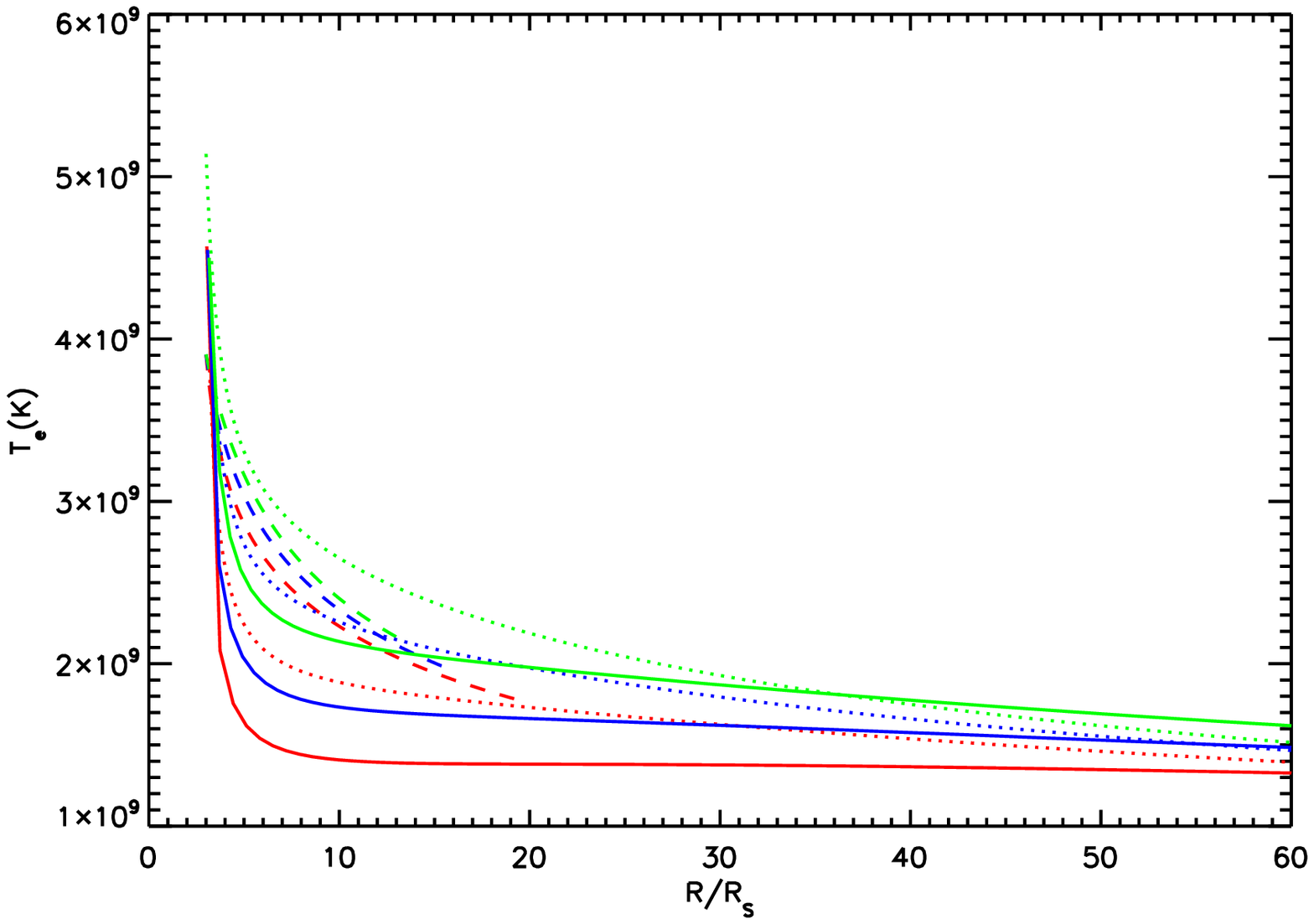}
\includegraphics[width=85mm,height=70mm,angle=0.0]{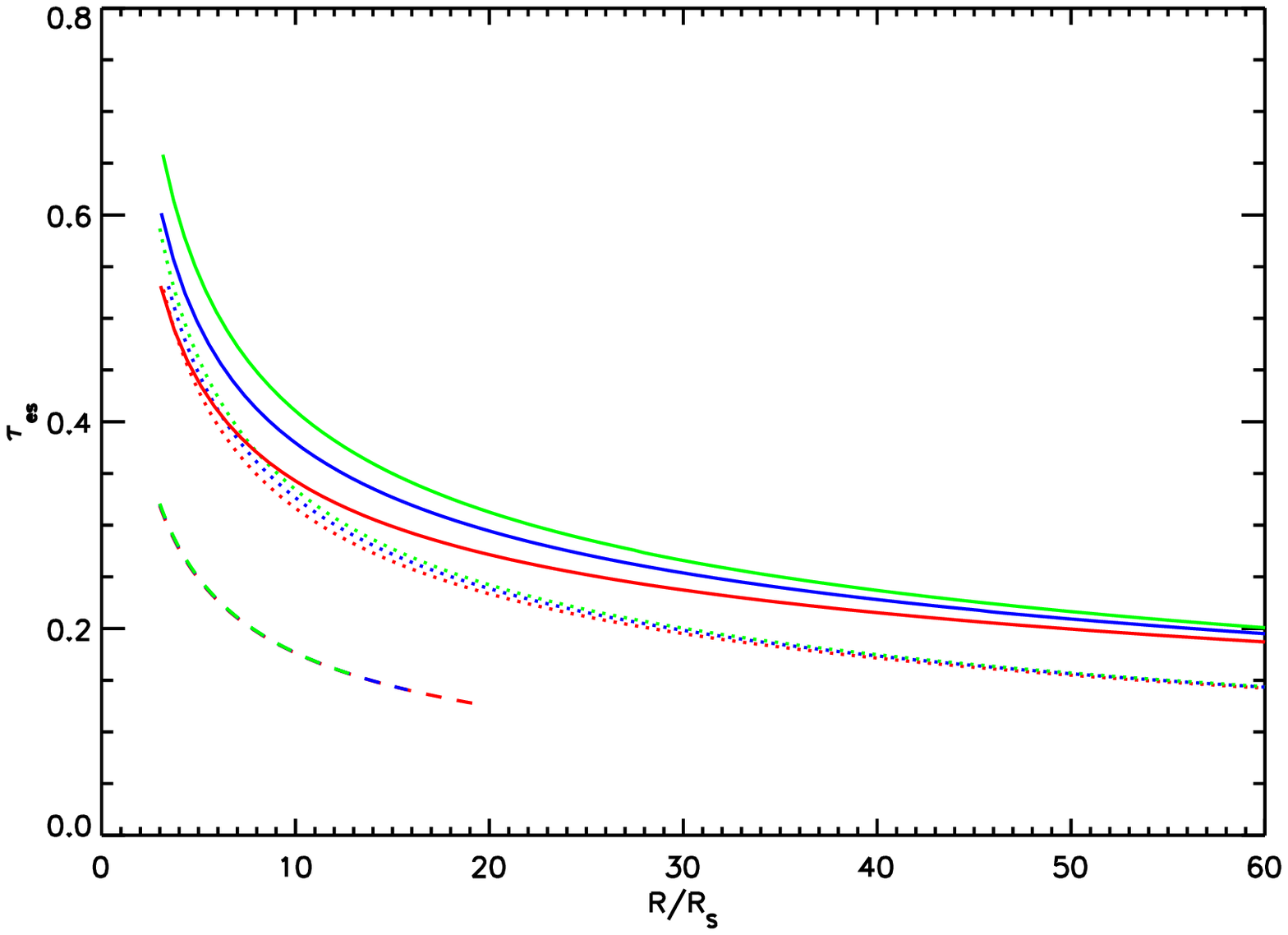}
\caption{\label{Te-tau}The left panel: the electron temperature of
the corona as a function of radius. In our calculation, $\alpha=0.2$
is adopted. The red lines represent the results calculated for
albedo $a=0.15$ (blue lines: $a=0.6$; green line: $a=1$). The
different line types represent different accretion rates (dashed
line: $\dot m=0.01$; dotted line $\dot m=0.02$; solid line: $\dot
m=0.03$). The right panel: the Thomson optical depth for the Compton
scattering of the hot corona in the vertical direction as a function
of radius.}
\end{figure*}

For viscosity parameter $\alpha=0.2$, and albedo $a=0.15$, it's
found that the inner disk can survive for accretion rate in the
range of $\dot m \sim 0.006-0.03$, which are consistent with the
results of Liu et al (2011), where the modification to the cooling
of the corona are not considered. This is because, at the upper
limit of the accretion rate, the corona is dominated by the Compton
cooling, the modification of conductive cooling to the corona can be
neglected, and at the lower limit of the accretion rate, the corona
is dominated by the Conductive cooling, the modification of Compton
cooling to the corona can be neglected. The size of the accretion
disk as functions of mass accretion rate is plotted in the left
panel of Fig. \ref{teff} with red line. The size of the inner disk
is defined by a critical radius, where neither the evaporation nor
the condensation precess occurs. Outside the critical radius, the
matter is evaporated to the corona from the disk (if the disk is not
evaporated completely) and inside the critical radius, the corona
matter condenses onto the disk. An increase of the mass accretion
rate leads to more gases condensation onto the disk and the
corresponding critical radius of the inner disk increases. The
maximum temperature of the inner disk is presented in the right
panel of Fig. \ref{teff} with red line. It's evident that the
maximum temperature of the disk increases with the mass accretion
rate. This is because, an increased value of the mass accretion rate
results in both the increase of the condensation rate and the
luminosity of the hot corona, the maximum temperature of the inner
disk heated up by accreting condensed gas and the irradiation of the
hot corona integrated throughout the inner disk increases. The ratio
of luminosity dissipated in the corona to the luminosity dissipated
in the disk, $L_{\rm c}/L_{\rm d}$, as functions of mass accretion
rates $\dot m$ is plotted in Fig. \ref{ratio} with red line. With
increase of the accretion rate $\dot m$, the relatively quick
increase of the emission from the accretion disk to the emission
from the corona leads to a decreased value of $L_{\rm c}/L_{\rm d}$.
The electron temperature and the optical depth for the Compton
scattering of the hot electron in the vertical direction as
functions of radius are plotted in Fig. \ref{Te-tau} with red line.
The solid line is for $\dot m=0.03$ (dotted line $\dot m=0.02$;
dashed line: $\dot m=0.01$). We can see the electron temperatures
are in the range of $\sim 1-5\times 10^{9}\rm \ K$ for different
accretion rate $0.01 \leq \dot m \leq 0.03$. It's clear that the
temperature of the corona decreases with increase of the mass
accretion rate for $a=0.15$. This is because, with increase of the
mass accretion rate, the Compton cooling of the soft photons from
the disk to the corona becomes more efficient, the temperature of
the corona decreases. To clearly show the mass distribution of the
inner disk-corona system in radial direction, we plot the mass
accretion rate in the accretion disk and the mass accretion rate in
the corona as functions of radius in Fig. \ref{mdot-corona}. It can
be seen that, with increase of the mass accretion rate, more gases
in the corona recondense back to the disk, the accretion rate in the
disk increases.

The emergent spectra of inner disk and corona with mass accretion
rate are plotted in Fig. \ref{spectrum1}. The X-ray spectrum in the
range of $2-10 \rm \ keV$ is roughly characterized by a power law
with a photon index $\Gamma=1.63$ at $\dot m=0.01$ (solid green
line). With increase of the mass accretion rate, the hard X-ray
photon index is $\Gamma=1.75$ at $\dot m=0.02$ (solid blue line) and
$\Gamma=1.92$ at $\dot m=0.03$ (solid red line). It's obviously that
the X-ray spectra become softer with increase of the mass accretion
rate. This can be understood as, the hard X-rays are dominated by
Bremsstrahlung at low accretion rate, whereas by Compton radiation
at high accretion rate. In Fig. \ref{spectrum1}, the dashed line is
the contribution of the Compton scattering of the soft photons from
the underlying cool disk by the electrons in the hot corona (Compton
radiation), and the dotted line is the contribution from the
bremsstrahlung of the transition layer. At lower accretion rate
$\dot m=0.01$, the cooling of the corona is dominated by the
conductive cooling, fewer soft photons from the underling thin disk
are scattered to the hard X-ray band, so the Compton emission to the
hard X-ray band is weak, the hard X-ray emission is dominated by the
bremsstrahlung of the transition layer. With increase of the mass
accretion rate to $\dot m=0.02$, the Compton emission of the corona
to the X-ray band becomes dominant compared to the bremsstrahlung of
the transition layer, as shown by the blue dashed line and dotted
line in Fig. \ref{spectrum1}.  With further increase of accretion
rate, the Compton $y$-parameter decreases, which results in a large
photon index and small luminosity ratio between the corona and disk,
as shown by the red solid line in Fig. \ref{spectrum1} and Fig.
\ref{ratio}.

In order to study the effect of albedo $a$ on the spectra, we plot
the emergent spectra of the inner disk and corona with albedo in
Fig. \ref{spectrum2}. The size and the maximum effective temperature
of the inner accretion disk with $a$ can also be seen in Fig
\ref{teff}. The electron temperature and the optical depth for the
Compton scattering of the hot electron in the vertical direction as
functions of radius are plotted in Fig. \ref{Te-tau}. The red line,
blue line and green line are for albedo $a=0.15$, $a=0.6$ and $a=1$
(without irradiation) respectively. For viscosity parameter
$\alpha=0.2$ and mass accretion rate $\dot m=0.02$, we plot the
emergent spectra for different albedo $a$ in Fig. \ref{spectrum2}.
For albedo $a=0.15$, the hard X-ray photon index $\Gamma=1.75$. With
increase the value of albedo, the hard X-ray photon index are
$\Gamma=1.63, 1.53$ for $a=0.6, 1$ respectively. It can be seen
clearly, X-ray spectrum becomes harder with increase of albedo. This
is because for larger albedo, fewer X-ray photons of the irradiation
from the corona to the disk are absorbed and reprocessed as thermal
seed photons to be scattered in the corona, the temperature of the
electron in the corona is higher, a harder X-ray spectrum is
predicted. We plot the emergent spectra with albedo $a$ for
different mass accretion rate in Fig. \ref{spectrum3}. We can see,
for both $\dot m=0.01$ and $\dot m=0.03$, the hard X-ray spectra
become harder with increase of $a$. From Fig. 6, it's also very
clear that the positive dependence of the hard X-ray spectrum index
on the mass accretion rate holds for different albedo $a$.

From theory, the value of albedo $a$ is uncertain, and has been
investigated by several authors, e.g. George \& Fabian (1991), White
et al. (1988). The value of albedo $a$, which depends on how the
incident photons from the corona interact with the underlying
accretion disk, is defined as the energy in the reflective spectrum
divided by the energy in the incident spectrum. The value of albedo
is energy-dependent. White et al. (1988) study the Compton
scattering of the incident photons by the electrons in the disk. In
their calculations, X-rays is assumed to incident upon a
semi-infinite, plane-parallel, zero temperature, purely scattering
medium. They show that the monochromatic albedo $a(x_{0})$ changes
from $95\%$ to $1\%$ for the energy $x_{0}$ in the range of $\sim
10^{-3}$ to $\sim 30$ ( $x_{0}$ is the energy of the incident
photons scaled with rest energy of electron). If the mean energy of
the photons are around 50 keV, $\sim 40 \%$ of the irradiation flux
is absorbed by the disk through the Compton scattering of the
electron in the disk (White et al. 1988). The K-shell absorption of
the metals is negligible for the incident photons at $\sim 50$ keV.
However, because the K-shell absorption of the metals is
proportional to $(m_{e}c^2/{h\nu})^{7/2}$, at lower energy, the
K-shell absorption becomes important. For the distribution of
$N_{\nu}\propto \nu^{-1.6}$, $\sim 30\%$ of the incident energy is
absorbed (Taam et al. 2008). Thus, $\sim 70\%$ of the total
irradiation flux is absorbed by the Compton scattering and K-shell
absorption, so the value of albedo $a$ is around 0.3. However, it's
highly uncertain for the calculation the value of albedo, e.g. the
ionization state of the surface of the accretion disk. In this
paper, we only take albedo as a parameter to fit the observations.

To test the effect of the viscosity parameters $\alpha$ to the shape
of the spectra, we plot the emergent spectra of the inner disk and
corona with viscosity parameter $\alpha$ in Fig. \ref{difalpha}, In
our calculation, albedo $a=0.15$ and mass accretion rate $\dot
m=0.03$ are adopted respectively. For $\alpha=0.2$, the hard X-ray
photons is $\Gamma=1.92$. With increase of $\alpha$, the hard X-ray
photon index are $\Gamma=1.67, 1.57$ for $\alpha=0.25, 0.3$
respectively. It's found that the luminosity of the disk-corona
system decreases systematically with the value of $\alpha$,
meanwhile, the X-ray spectra in the range of 2-10 keV becomes harder
with the value of $\alpha$. This is because, for fixed mass
accretion rate, the surface density in the corona decreases with
increase of $\alpha$. With a lower density, the heat flux to the
transition layer decreases, meanwhile the bremsstrahlung cooling
rate is decreased even more. The energy balance between heating and
cooling results in a decreased heating rate associated with a lower
enthalpy flux. The net effect results in a decreased condensation
rate. The energy conversion efficiency of the inner disk +
corona/ADAF can be expressed as $\eta/{0.1}={(L/L_{\rm Edd})} /
{(\dot M/ \dot M_{\rm Edd})}$. The efficiency of the thin disk is
0.1, however, for the hot corona/ADAF, due to the strong advection
effect, the efficiency is much less than 0.1. With increase of
$\alpha$, due to the decrease of the condensation rate, less matters
are accreted in the form of thin disk, the luminosity of the inner
disk + corona/ADAF system decreases. Meanwhile, due to relative lack
photons from the underlying thin disk to be scattered in the corona,
the temperature of the corona is higher, so a harder X-ray spectrum
is predicted.

To show the effect of the different parameters (accretion rate $\dot
m$, viscosity parameter $\alpha$ and albedo $a$) to the condensation
features and the corresponding spectral features of the inner disk
and corona clearly, we list the numerical results for different
parameters in Table 1.

\section{Comparison To Observations}
\subsection{\rm GX 339-4}
The distance to GX 339-4 is dynamically constrained at $6.7 < d <9.3
\rm \ kpc$ (Zdziarski et al. 2004), which is consistent with the
lower limit of $d\gtrsim 6\rm \ kpc$ for kinematical distance based
on analysis of high-resolution spectra of the Na D line (Hynes et
al. 2004). The mass function of GX 339-4 is constrained to $\sim
6M_{\odot}$ (Hynes et al. 2003; Mu$\rm \tilde{n}$oz-Darias et al.
2008). In this paper, a likely distance of $d=8\rm \ kpc$  and mass
$m=5.8$ are adopted to fit the recent observations (Tomsick et al.
2008). A thermal component with an inner disk temperature of
$0.165\rm \ keV$ at a lower luminosity $L\approx0.008 L_{\rm Edd}$
(1-100\ keV) are detected by the observations of {\em Swift} and
{\em RXTE} during 2007 June 10-14. The most constraints of the disk
come from the reflection component, which requires that the disk can
extend to $ r < 5 $ (Tomsick et al. 2008).

With $m=5.8$, the X-ray luminosity $L=0.008L_{\rm Edd}$ and
effective temperature of the disk $T_{\rm eff, max}=0.165$ from
observations. We assume viscosity parameter $\alpha=0.3$, and find
that the observed X-ray luminosity and the effective temperature are
matched by adopting $\dot m=0.0456$ and albedo $a=0.782$
respectively. The fitting results are listed in Table 2. We
calculate the emergent spectra of the inner disk-corona system with
the fitting parameters, and plot the spectrum in Fig. \ref{gx339-4}
(the solid line). It's found that the hard X-ray spectrum in the
range of $1-100 \rm \ keV$ is roughly characterized by a power law
with a photon index $\Gamma_{\rm mod}=1.50$ from our model, which is
roughly close to the observed one $\Gamma=1.63^{+0.04}_{-0.03}$
(Tomsick et al. 2008).

With increase of the viscosity parameter to $\alpha=0.4$, we find
the observed X-ray luminosity and the maximum effective temperature
of the disk can be matched by taking accretion rate $\dot m=0.0773$
and albedo $a=0.617$ respectively. In this case, the condensation
rate is $\dot m_{\rm cnd}=5.98\times 10^{-4}$ which is smaller than
$\dot m_{\rm cnd}=1.4\times 10^{-3}$ for the case $\alpha=0.3$. So
from eqs. (\ref{tmaxp}) and (\ref{tmax}), a smaller value of albedo
$a$ is needed to match the observed effective temperature. We plot
the corresponding emergent spectrum in Fig \ref{gx339-4} (the dotted
line). The hard X-ray spectrum from $1-100 \rm \ keV$ is also
roughly characterized by a power law with a photon index
$\Gamma_{\rm mod}=1.4$ from our model. The detailed fitting results
are shown in Table 2.

From the calculation above, it's found that, the hard X-ray photon
index calculated by taking $\alpha=0.3$ can match the observed one
better than that of taking $\alpha=0.4$. Here, we want to show that
the observed X-ray luminosity and the maximum effective temperature
of the disk can only be fitted in a very narrow range for the value
of $\alpha$. For instance, for $\alpha=0.2$, we can't fit the X-ray
luminosity and the maximum effective temperature simultaneously,
i.e., by taking $\dot m=0.0251$, the observed X-ray luminosity can
be matched, however, even by taking the limited value of albedo
$a=1$, the theoretical value of $T_{\rm eff, max }$ is $0.1825 \rm \
keV$ which is bigger than the observed value $0.165 \rm \ keV$.

\subsection{\rm Cyg X-1}
Cyg X-1 is a well-known black hole X-ray binaries and has been well
studied by several authors. The distance to Cyg X-1 is very early
estimated at $2\ \rm kpc$ (Murdin \& Webster 1971; Reis et al.
2010), and confirmed by Massey et al. (1995) where a distance of
$2.1\pm 0.1\rm \ kpc$ is found. The mass of Cyg X-1 is found in the
range of $7-15\ M_{\odot}$ (Shaposhnikov \& Titarchuk 2009). In this
paper, a distance $d=2 \rm \ kpc$ and $m=10$ are adopted (Reis et
al. 2010). We compare our result with the recent observations of Cyg
X-1 (Reis et al. 2010). They find that an X-ray luminosity $L_{\rm
x}/L_{\rm Edd}=5\times10^{-3}(m/10)\times(d/2\rm \ kpc)^{2}$ in the
range of $0.5-10 \rm \ keV$ and a hard power-law photon index
$\Gamma=1.71\pm 0.01$ (2-10\ keV). Extrapolating the luminosity to
$100 \rm \ keV$, we get a X-ray luminosity $L_{\rm x}/L_{\rm
Edd}=1.47\times 10^{-2}(m/10)\times(d/2\rm \ kpc)^{2}$. By fitting
the spectra with Diskbb+Power law model, they reveal a cool disk
component with a temperature $kT\approx 0.194\rm keV$, meanwhile
it's found that the inner boundary of the accretion disk is very
close to ISCO, i.e., $r_{\rm in}=5.7^{+7.0}_{3.0}$ (Reis et al.
2010).

With $m=10$, the X-ray luminosity  $L_{\rm x}=1.47\times
10^{-2}L_{\rm Edd}$ and effective temperature of the disk $kT\approx
0.194\rm keV$ derived from observations, and assuming viscosity
parameter $\alpha=0.3$, it's found that the observed X-ray
luminosity and effective temperature are matched by taking $\dot
m=0.044$ and albedo $a=0.398$ respectively. We plot the emergent
spectrum of the inner disk and corona with the fitting parameters in
Fig. \ref{cygx-1} (the solid line). It's found that the hard X-ray
spectra in the range of $2-10 \rm \ keV$ is characterized by a power
law with a photos index $\Gamma_{\rm mod}=1.60$ from our model. The
fitting results for bigger viscosity parameter $\alpha=0.4$ are also
listed in Table. 2.

\section{Conclusion}
We study the formation of a cool, optically thick, geometrically
thin disk in the most inner region of black hole X-ray binaries in
the low/hard spectral state within the context of disk accretion fed
by condensation of hot corona. By taking into account the cooling
process associated with both Compton scattering and vertical
conduction in a corona, and the irradiation of the hot corona to the
disk, we obtain a self-consistent solution of the inner disk and
corona. We calculate the emergent spectra of the inner disk and
corona for different accretion rate, and examine the effect of
different parameters on the spectra (e.g. viscosity parameters
$\alpha$ and albedo $a$). Our model is used to explain the spectral
features of BHXRBs GX 339-4 and Cyg X-1 in the low/hard spectral
state, in which the thin disk resides in the region very closed to
the ISCO.  Our results are roughly in agreement with the
observations characterized by a weak disk component with temperature
of $\sim 0.2$ keV and a hard power-law X-ray spectra with $\Gamma
\sim 1.5-2.1$ for the sources in the low/hard spectral state.

\acknowledgments
 We thank Prof. Ronald E. Taam for his very useful
suggestions and comments. We thank the referee for his/her expert
comments and suggestions to our paper. This work is supported by the
the National Natural Science Foundation of China (grants 11033007,
11173029) and by the National Basic Research Program of China-973
Program 2009CB824800.

\begin{figure*}\label{mdot-corona}
\includegraphics[width=85mm,height=70mm,angle=0.0]{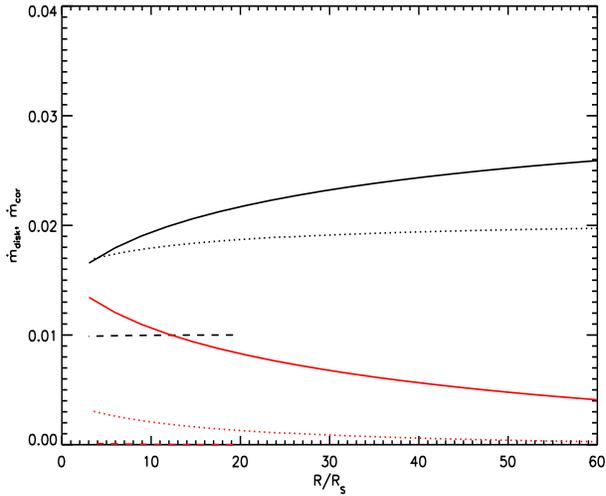}
\caption{\label{mdot-corona} The mass accretion rate in the
accretion disk and the mass accretion rate in the corona as
functions of radius. The red line and black line represent the mass
accretion rate in the disk and the mass accretion rate in the corona
respectively. The solid line, dotted line and dashed line are for
$\dot m=0.03$, $\dot m=0.02$ and $\dot m=0.01$ respectively. In our
calculation, $\alpha=0.2$ and albedo $a=0.15$ are adopted.}
\end{figure*}

\begin{figure*}
\includegraphics[width=85mm,height=70mm,angle=0.0]{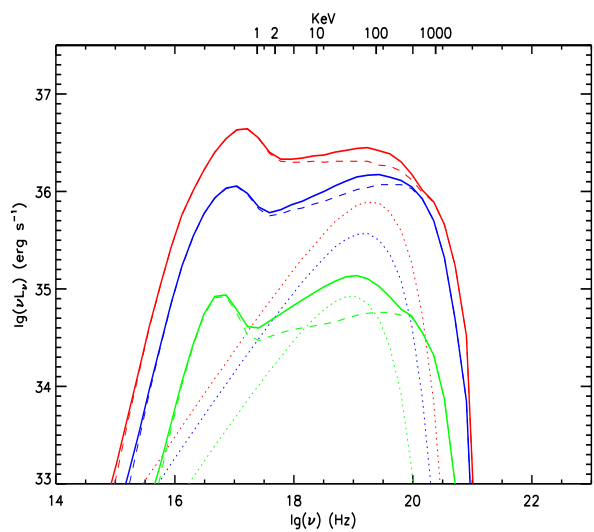}
\caption{\label{spectrum1}The emergent spectra of inner disk and
corona for different mass accretion rate. In our calculation,
$\alpha=0.2$ and albedo $a=0.15$ are adopted. The red line,  blue
line and green line are the spectra for $\dot m=0.03$,  $\dot
m=0.02$ and $\dot m=0.01$ respectively. The solid line is the total
spectrum, the dashed line is the contribution from the
Comptonization of the soft photons of the thin disk by the electrons
in the hot corona, the dotted line is the contribution of the
bremsstrahlung from the transition layer.}
\end{figure*}

\begin{figure*}
\includegraphics[width=85mm,height=70mm,angle=0.0]{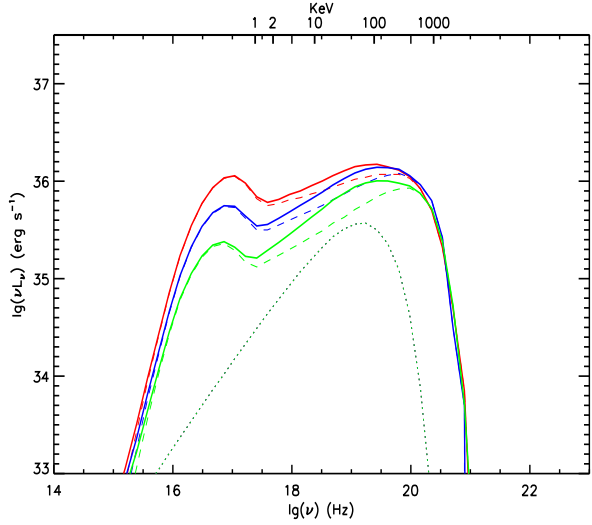}
\caption{\label{spectrum2}The emergent spectra of inner disk and
corona for different albedo $a$. In our calculation, $ \dot m=0.02$
and $\alpha=0.2$ are adopted respectively. The solid red line is the
total spectrum for $a=0.15$, the red dashed line is the contribution
from the Comptonization of the soft photons of the thin disk by the
hot corona, the red dotted line is the contribution of the
bremsstrahlung from the transition layer. The solid blue line and
green line are for $a=0.6$ and $a=1$ respectively.}
\end{figure*}

\begin{figure*}
\includegraphics[width=85mm,height=70mm,angle=0.0]{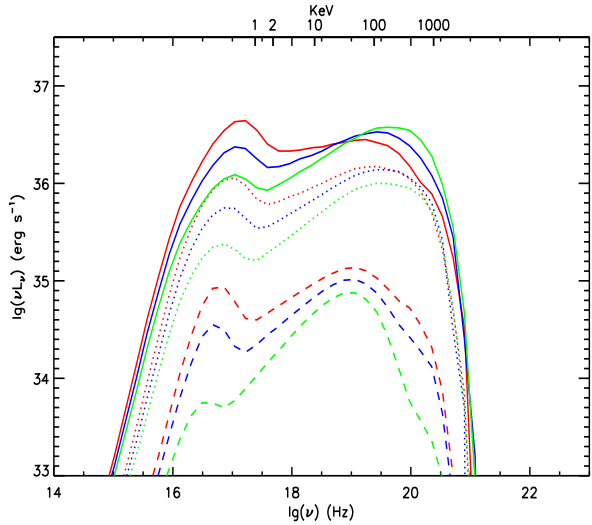}
\caption{\label{spectrum3}The emergent spectra of inner disk and
corona for different accretion rate and albedo.  In our calculation,
$\alpha=0.2$ is adopted. The solid line, dotted line and dashed line
are for $\dot m=0.03$, $\dot m=0.02$, and $\dot m=0.01$
respectively. The red line, blue line and green line are for albedo
$a=0.15$, $a=0.6$ and $a=1$ respectively.}
\end{figure*}

\begin{figure*}
\includegraphics[width=85mm,height=70mm,angle=0.0]{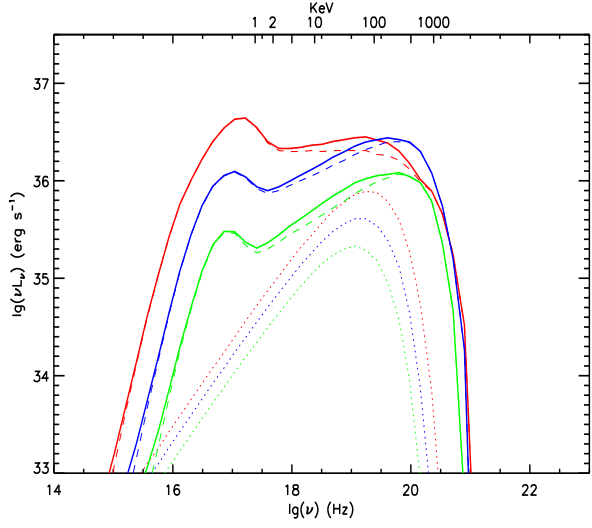}
\caption{\label{difalpha}The emergent spectra of inner disk and
corona for different viscosity parameter $\alpha$. In our
calculation, $\dot m=0.03$ and $a=0.15$ are adopted. The red solid
line is the total spectrum for $\alpha=0.2$, the red dashed line is
the contribution from the Comptonization of the soft photons of the
thin disk by the hot corona, the red dotted line is the contribution
of the bremsstrahlung from the transition layer. The solid blue line
and green line are for $\alpha=0.25$ and $\alpha=0.3$ respectively.}
\end{figure*}

\begin{table*}\label{T:result}
\begin{center}
\begin{minipage}{\linewidth}
\caption{Condensation and spectral features of the inner disk and
corona around the black hole of 10 $M_{\odot}$ }
\begin{tabular}{ccc|cccccc}
\hline\hline
&&&&&&& \\
$\alpha$&  $a$  &$\dot m$  & $r_{\rm d}$ & $\dot m_{\rm cnd}$ &
$L_{\rm x(cor)}/L_{\rm Edd}$ & $L_{\rm c}/L_{\rm d}$ &
$T_{\rm eff,max} (\rm keV)$ & $\Gamma_{\rm mod}$ \\
&&&&&&&& \\
\hline
&&&&&&&& \\
0.2             & 0.15& 0.01 & 19.2 &  $9.66\times10^{-5}$ & $1.02\times10^{-3}$ &  $3.04$ &0.10& 1.63\\
0.2             & 0.15& 0.02 & 124.2 & $2.98\times10^{-3}$ & $9.18\times10^{-3}$ &  $1.56$ &0.19& 1.75\\
0.2             & 0.15& 0.03 & 350.0 & $1.29\times10^{-2}$ & $3.32\times10^{-2}$ &  $1.45$ &0.27& 1.92\\
\hline
&&&&&&&& \\
0.2             & 0.6 & 0.01  & 15.8  & $5.92\times10^{-5}$   & $8.22\times10^{-4}$ &   $6.50$  &0.08 & 1.52\\
0.2             & 1   & 0.01  & 13.1  & $3.60\times10^{-5}$   & $6.79\times10^{-4}$ &   $43.0$  &0.05 & 1.40\\
0.2             & 0.6 & 0.02  & 106.2  & $2.25\times10^{-3}$   & $6.71\times10^{-3}$ &  $2.27$  &0.16 & 1.63\\
0.2             & 1   & 0.02  & 92.9  & $1.66\times10^{-3}$   & $5.05\times10^{-3}$ &   $4.06$  &0.13 & 1.53\\
0.2             & 0.6 & 0.03  & 308.2  & $1.07\times10^{-2}$   & $2.48\times10^{-2}$ &  $1.93$  &0.23 & 1.77\\
0.2             & 1   & 0.03  & 278.3  & $8.72\times10^{-3}$   & $1.83\times10^{-2}$ &  $2.72$  &0.20 & 1.62\\
\hline
&&&&&&&& \\
0.2             & 0.15 & 0.03  & 350.0  & $1.29\times10^{-2}$   & $3.32\times10^{-2}$ &  $1.45$ &0.27  & 1.92\\
0.25            & 0.15 & 0.03  & 83.0  & $3.19\times10^{-3}$   & $1.35\times10^{-2}$  &  $1.79$ &0.21  & 1.67\\
0.3             & 0.15 & 0.03  & 22.2  & $4.66\times10^{-4}$   & $4.25\times10^{-3}$  &  $2.80$  &0.15 & 1.57\\

\hline
\end{tabular}
\\
\\
Note.--- With black hole mass $m$=$10$, viscosity parameters
$\alpha$, albedo $a$, and the mass accretion rate $\dot m$, the size
of the inner disk $r_{\rm d}$, condensation rate $\dot m_{\rm cnd}$
integrated from the condensation radius to $3 R_{\rm S}$, the
luminosity dissipated in the corona $L_{\rm x(cor)}/L_{\rm Edd}$,
the ratio of luminosity dissipated in the corona to the luminosity
dissipated in the disk $L_{\rm c}/L_{\rm d}$, the maximum
temperature of the inner disk $T_{\rm eff,max}$ and the hard X-ray
photon index $\Gamma_{\rm mod}$ in the range of 2-10 keV are listed.
\end{minipage}
\end{center}
\end{table*}


\begin{figure*}
\includegraphics[width=85mm,height=70mm,angle=0.0]{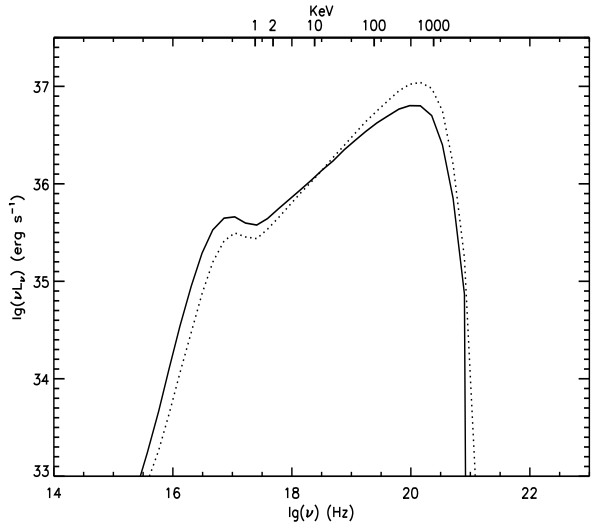}
\caption{\label{gx339-4}The emergent spectra of GX 339-4 calculated
by different fitting parameters of our model. The solid line is the
emergent spectrum for $m=5.8$, $\alpha=0.3$, $\dot m=0.0456$ and
$a=0.782$; The dotted line is the emergent spectrum for $m=5.8$,
$\alpha=0.4$, $\dot m=0.0773$ and $a=0.617$.}
\end{figure*}

\begin{figure*}
\includegraphics[width=85mm,height=70mm,angle=0.0]{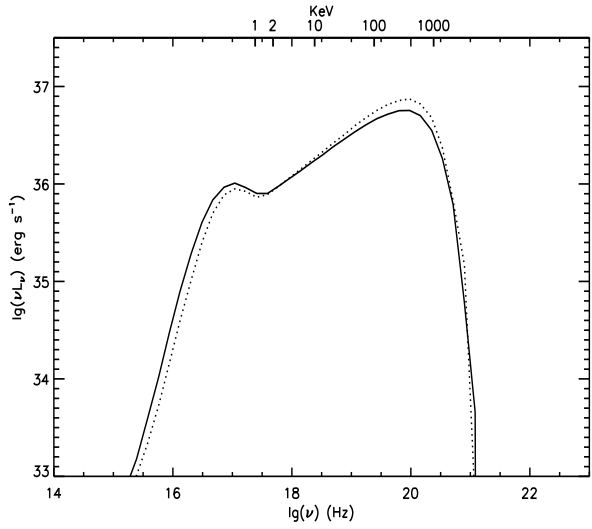}
\caption{\label{cygx-1}The emergent spectra of Cyg X-1 calculated by
different fitting parameters of our model. The solid line is the
emergent spectrum for $m=10$, $\alpha=0.3$, $\dot m=0.044$ and
$a=0.398$; The dotted line is the emergent spectrum for $m=10$,
$\alpha=0.4$, $\dot m=0.064$ and $a=0.267$.}
\end{figure*}

\begin{table*}\label{T:fit}
\begin{center}
\begin{minipage}{\linewidth}
\caption{Fitting Results for GX 339-4 and Cyg X-1}
\begin{tabular}{cccccccc}
\hline\hline
&&&&&& \\
\multicolumn{8}{c}{GX 339-4:  ($m=5.8$$^{a}$, $L_{\rm x}/L_{\rm
Edd}=0.008$$^{b}$, \ $T_{\rm eff, max}=0.165\ \rm keV$$^{c}$,
$\Gamma=1.63^{+0.04}_{-0.03}$$^{d}$)}\\
&&&&&&& \\
\hline
$\alpha$&  $\dot m$  &$a$  & $r_{\rm d}$ & $\dot m_{\rm cnd}$ & $L_{\rm x(cor)}/L_{\rm Edd}$ & $L_{\rm c}/L_{\rm d}$ & $\Gamma_{\rm mod}$ \\
\hline
0.3.......................& 0.0456 & 0.782 & 34.9 & $1.40\times10^{-3}$ & $0.8\times10^{-2}$ &  $5.0$ & 1.50\\
0.4.......................& 0.0773 & 0.617 & 16.05& $5.98\times10^{-4}$ & $0.8\times10^{-2}$  &  $6.59$ & 1.40\\
\hline
&&&&&&& \\
\multicolumn{8}{c}{Cyg X-1: ($m=10$$^{a}$,\ $L_{\rm x}/L_{\rm
Edd}=0.0147$$^{b}$,\ $T_{\rm eff, max}=0.194$\ keV$^{c}$,\ $\Gamma=1.71\pm 0.01$$^{d}$)} \\
&&&&&&& \\
\hline
$\alpha$&  $\dot m$  &$a$  & $r_{\rm d}$ & $\dot m_{\rm cnd}$ & $L_{\rm x(cor)}/L_{\rm Edd}$ & $L_{\rm c}/L_{\rm d}$ & $\Gamma_{\rm mod}$ \\
\hline
0.3.......................& 0.044 & 0.398 &55.0    &  $2.73\times10^{-3}$  & $1.47\times10^{-2}$ &  $2.57$  & 1.60\\
0.4.......................& 0.064 & 0.267 & 30.4   &  $1.55\times10^{-3}$  & $1.47\times10^{-2}$ &  $2.89$  & 1.52\\
\hline
\end{tabular}
\\
\\
Note.---With the black hole mass $m$, the X-ray luminosity $L_{\rm
x(cor)}/L_{\rm Edd}$ and the effective temperatur $T_{\rm eff, max}$
of the accretion disk derived from observation, and assuming a value
of $\alpha$, two quantities accretion rate $\dot m$ and albedo $a$
are adjusted simultaneously to match the observed X-ray luminosity
and the effective temperatur of the accretion disk. $r_{\rm d}$,
$\dot m_{\rm cnd}$, $L_{\rm x(cor)}/L_{\rm Edd}$ and $L_{\rm
c}/L_{\rm d}$ are the corresponding size of the inner disk, the
integrated condensation rate from the condensation radius to
$3R_{\rm S}$, the ratio of luminosity dissipated in the corona to
the luminosity dissipated in the disk respectively. $\Gamma_{\rm
mod}$ is the hard X-ray
photon index in the range of $2-10\rm \ keV$ calculated by the corresponding parameters.\\
\footnotemark{$a$}The central black hole mass\\
\footnotemark{$b$}The observed X-ray luminosity\\
\footnotemark{$c$}The observed temperature of the accretion disk\\
\footnotemark{$d$}The observed X-ray photon index in the range of
2-10\ keV
\end{minipage}
\end{center}
\end{table*}
\end{document}